\documentclass[conference]{IEEEtran}
\usepackage{array}
\usepackage{graphicx}
\usepackage{amsmath}
\usepackage{algorithmic}
\usepackage{algorithm}
\usepackage{epsfig}
\usepackage{url}
\usepackage{epstopdf}
\usepackage{cite}
\usepackage{setspace}
\usepackage{enumitem}
\usepackage{graphicx,lipsum}
\usepackage{caption}
\usepackage[caption=false]{subfig}
\usepackage{tikz}

\newcolumntype{P}[1]{>{\centering\arraybackslash}p{#1}} 
\newcolumntype{M}[1]{>{\centering\arraybackslash}m{#1}} 
\newcommand\copyrighttext{%
  \footnotesize \copyright~2015 IEEE. Personal use of this material is permitted. Permission from IEEE must be obtained for all other uses, in any current or future media, including reprinting/republishing this material for advertising or promotional purposes, creating new collective works, for resale or redistribution to servers or lists, or reuse of any copyrighted component of this work in other works.}
\newcommand\copyrightnotice{%
\begin{tikzpicture}[remember picture,overlay]
\node[anchor=south,yshift=10pt] at (current page.south) {\fbox{\parbox{\dimexpr\textwidth-\fboxsep-\fboxrule\relax}{\copyrighttext}}};
\end{tikzpicture}%
}
\begin{document}
\bibliographystyle{IEEEtran}
%

\title{Evaluation of Channel Assignment Performance Prediction Techniques in Random Wireless Mesh Networks}

\author{\IEEEauthorblockN{M Pavan Kumar Reddy, Srikant Manas Kala and Bheemarjuna Reddy Tamma}
\IEEEauthorblockA{ Indian Institute of Technology Hyderabad, India\\
Email: [cs12b1025, cs12m1012, tbr]@iith.ac.in}}

\maketitle
 \copyrightnotice
\begin{abstract}

Performance of wireless mesh networks (WMNs) in terms of network capacity, end-to-end latency, and network resilience depends upon the prevalent levels of interference. Thus, interference alleviation is a fundamental design concern in multi-radio multi-channel (MRMC) WMNs, and is achieved through a judicious channel assignment (CA) to the radios in a WMN. In our earlier works we have tried to address the problem of estimating the intensity of interference in a wireless network and predicting the performance of CA schemes based on the measure of the interference estimate. We have proposed reliable CA performance prediction approaches which have proven effective in grid WMNs. In this work, we further assess the reliability of these CA performance prediction techniques in a large MRMC WMN which comprises of randomly placed mesh routers. We perform exhaustive simulations on an ns-3 802.11n environment. We obtain conclusive results to demonstrate the efficacy of proposed schemes in random WMNs as well. 
\end{abstract}

\section{Introduction}
Wireless mesh network (WMN) deployments are traditionally designed in grid or grid-like structures, most common being the square grids, triangular grids and the hexagonal grids. The reason behind the choice of a grid topology is the significant enhancement in network capacity that is guaranteed by a grid WMN, which is shown to be almost double of the network capacity of a random WMN (RWMN) comprising of same number of mesh routers (MRs) \cite{Grid}. Similarly, RWMNs are worse off than grid WMNs in terms of network span as well. Gateway placement strategies also rely on grid WMNs to harness optimal aggregate throughput \cite{Grid3}.\\
Despite the benefits that grid WMN layouts offer when compared to arbitrary WMN layouts, there are certain application scenarios where random WMNs with strategic placement of MRs is required. Two very important applications of random wireless networks are in the \textit{rural information networks} (RINs) and \textit{agricultural wireless sensor networks} (AWSNs). These two domains are highly relevant in the developing nations to offer their rural population improved amenities, better administration, greater access to information and prompt weather and disaster alerts through reliable Internet connectivity. Rural WMNs hold great potential in India as 70\% of its population lives in rural areas and 58\% of its rural households primarily depend on agriculture sector for livelihood \cite{census}. It is evident from the fact that the Govt. of India has launched a \textit{Rurban Mission} (Rural-Urban Mission) which primarily focuses on providing urban facilities in rural areas and depends heavily on RINs and AWSNs.

\section{Motivation and Related Research Work}

Interference is highly undesirable in wireless networks and there is a constant effort in academic and industrial research to restrict its impact on the network performance. An indispensable tool in the task of interference mitigation is a \textit{Conflict graph} (CG). It does a comprehensive accounting of the endemic interference scenarios and delivers a reliable interference estimate of a WMN. This estimate is called \textit{interference degree} (ID) at the granularity of a link and \textit{total interference degree} when computed for the entire WMN. ID is the measure of potential conflicts that transmission on a link may be hampered by while TID is the sum of all such potential transmission conflicts in the WMN \cite{22Ramachandran}. It is discernible that a high magnitude of the TID estimate signifies high interference levels in the network and vice-versa.

In \cite{Manas3} we propose a new approach to characterize interference, wherein we split the interference scenarios prevalent in a wireless network into three categories \textit{viz.,} statistical, spatial and temporal. We utilize this 3 dimensional model to devise a statistical interference metric based on the \textit{Channel Distribution Across Links} approach named CDAL$_{cost}$. It is computed by first determining the channel(s) assigned to each link in the network and then determining the total number of links assigned to each channel. CDAL$_{cost}$ is a measure of equitable distribution of channels across radios in a wireless network and is based on the notion of fairness in channel utilization. Thus, a CA with an even distribution and utilization of channels has a lower CDAL$_{cost}$. Like TID, magnitude of CDAL$_{cost}$ too is inversely proportional to CA performance. Further, in \cite{Manas4} we demonstrate that a statistical interference metric such as CDAL$_{cost}$ will fail to account for the 
spatial features of interference. We then propose a spatio-statistical approach predicated on the 3 dimensional interference characterization model of \cite{Manas3} which considers a link level view of a wireless network. It considers a set of $X$ links as the fundamental unit for which a \textit{weight} is computed. This weight reflects the degree of adverse impact of interference on the network. The cumulative weight of all such sets of $X$ links in the entire WMN offers the \textit{Cumulative X-Link-Set Weight} or $CXLS_{wt}$ of the network. The parameter $X$ is of great significance, as it denotes the \textit{interference impact radius} of a particular link. $X$ is derived by the ratio of interference range (I$_r$) and transmission range (T$_r$) of a radio \emph{i.e.,} $X = I_r:T_r$. For ease of implementation, $X$ is assumed to be a positive integer. 

In \cite{Manas3} and \cite{Manas4}, we demonstrate through rigorous simulations that $CXLS_{wt}$ is the most reliable interference and CA performance prediction metric among the three. Further, CDAL$_{cost}$ decidedly outperforms TID and delivers better CA performance predictions. The network parameters used to assess the reliability of each of these metrics were network capacity, end-to-end latency and packet loss ratio.\\
However, in both the works the topology of the simulation set-up was limited to a grid WMN. The efficacy and performance of these schemes was not tested in a WMN where MRs are randomly placed and do not adhere to a specific pattern. This leaves a room for doubt in the concreteness of the results as the test set-up was not comprehensive. In this work, we fill that gap by carrying out extensive simulations in a large scale WMN with randomly placed MRs \emph{i.e.,} an RWMN.

\section{Relevance of a Random WMN}
The deployment constraints of RWMNs, which primarily include RINs and AWSNs, substantially differ than those of urban or enterprise WMN deployments. We list the important design constraints below and demonstrate the inability of grid WMN layouts to offer a practical solution.
\begin{itemize}
 \item \textbf{Resource Limitations:~} Constraints which are primarily financial, and up to some extent logistical and technical, render RINs and AWSNs with a resource crunch especially in the form of limited availability of IEEE 802.11 radios. Thus a grid WMN which requires a large number of MRs to be located in close proximity in a fixed spatial pattern is not a viable option in the rural deployments. 
 \item \textbf{Connectivity Constraints:~} Rural areas are a few steps behind urban centers with respect to availability of connectivity technologies. The absence of all Internet connectivity is a norm rather than the exception, and can only be addressed through a satellite connection or a long range IEEE 802.11 based connection that employs highly directional antennas. In some rural regions, government administrative centers may be equipped with a dedicated broadband connection that can be harnessed to provide local wireless connectivity through off-the-shelf 802.11 commodity hardware \cite{GridRural3}. Connectivity constraints have a two-fold impact on rural WMN deployments. First, satellite or long-range 802.11 connections raise the fixed and operational costs of networks rendering them economically less viable. Second, the services offered through such RINs/AWSNs \emph{viz.}, a basic interactive/transactional service, a high bandwith data service or both, depend on the connectivity technology being 
employed. 
A prudent placement of MRs becomes very crucial in such deployments to mitigate cost escalation and enhance the QoS provided by the network. Further, the connectivity exhibited by wireless networks with a selective deployment of MRs is comparable to grid topology WMNs when networks of similar densities are observed \cite{Grid}.    
 \item \textbf{Geographical Limitations:~} The most challenging design constraint is the optimal placement of nodes owing to the rural terrain. Strategic locations need to be identified for placement of MRs so as to maximize signal penetration across the target area and avoid any \textit{blind-spots} that can be created by obstacles peculiar of rural geography, \emph{e.g.}, tall trees, hills etc. 
 A poorly planned RIN/AWSN will suffer from weak signal intensity, high end-to-end delays and inconsistent connectivity \cite{GridRural2}. The need to successfully meet these criteria in rural deployments would also tilt the debate in favor of RWMNs, as they offer a flexibility in node placement that grid layouts do not.  
  \item \textbf{Data Collection:~} Most rural networks are designed to seamlessly collect and disseminate data, which requires some MRs to function as local data centers. These central nodes need a robust and sustained connectivity to the gateway nodes, with some redundancy in data paths for enhanced fault resilience. In addition, an \textit{intra-network} control and monitoring mechanism for operational purposes is also necessary \cite{GridRural}. Grid layouts are rigid in deployment and will be unable to implement an information system that prioritizes a few nodes and offers them favorable access to the Gateway. 
   \item \textbf{Maintenance Overhead:~} As the complexity of a system rises, so does the operational overhead. Maintenance costs in rural WMNs have an additional cost component of physical fault inspections and diagnosis which includes the high transportation expenses incurred in reaching the rural centers. A complex grid WMN with high fault-tolerance and redundancy does offer a reliable communication system, but it also substantially scales up operational costs that are not feasible for deployments in rural centers.
\end{itemize}

Taking all these factors into consideration, it is easy to infer that RWMNs are the deployments of choice for remote WMN deployments such as RINs and AWSNs. They are a class of WMNs that is highly relevant from both, a technological and a socio-economic perspective, and need to be thoroughly studied in terms of network performance. Thus, in this study we consider a RWMN deployment and subject the efficacy of our CA performance prediction techniques. 
\section{Simulations, Results and Analysis}
\subsection{Topological Considerations of the RWMN}
Graphs are widely used to model complex communication networks. These representations exhibit features which are typical of the topology of the communication networks they attempt to model. The model of a real-world communication network would result in a complex graph that is neither regular nor entirely arbitrary \cite{Complex}. In order to establish closeness to real-world networks in our RWMN deployment, we design a wireless network that is not purely arbitrary or random but adheres to topological parameters of a real-world network. Topological metrics are local \emph{i.e.}, node-centric, and global \emph{i.e.}, graph centric. Parameters such as \textit{Degree}, \textit{Eccentricity}, \textit{Betweenness Centrality} are local parameters while \textit{Density} and \textit{Transitivity} are global measures of a graph. In the RWMN design process, we focus on emulating global measures of real-world wireless networks.

We design a RWMN of 50 nodes which spans a simulated environment of $1500m \times 1500m$. The links between nodes (MRs) in the RWMN, represented by edges between the corresponding vertices, are added in such a way that the following global parameters of the RWMN (G$_r$) adhere to corresponding real-world values \cite{TOPO}. 
\begin{itemize}
 \item \textbf{Density ($\delta(G_r)$) :~} It is the ratio of wireless links that will be used for communication in the RWMN and the total number of links that can be established \emph{i.e.}, $\textsuperscript {50} C_2$. ($\delta$) of G$_r$ is computed as $l/\textsuperscript n C_2$, where $n$ is the number of nodes and $l$ is the number of active wireless links.
 \item \textbf{Diameter ($D(G_r)$) :~} It is the maximal \textit{eccentricity} exhibited by the network \emph{i.e.}, the maximal geometric distance between any two nodes in  G$_r$.
 \item \textbf{Radius ($R(G_r)$) :~} It corresponds to the minimal eccentricity exhibited by the network.
 \item \textbf{Clustering Coefficient ($CC(G_r)$) :~} It denotes the degree to which nodes in a network form closely knit clusters. It is based on the concept of triplets of nodes, and is the ratio between the number of \textit{closed triplets} or triangles formed by the nodes in the graph, upon the total number of possible triplets in the graph.
\end{itemize}

We present the range of these global measures, along with the range of values exhibited by real-world networks and the corresponding values of the RWMN being designed in Table~\ref{comp}. It can be inferred that $\delta(G_r)$ of the considered RWMN lies slightly below the prescribed range for real-world networks. The reason for this consideration is the fact that rural WMNs are less likely to be as dense as urban networks and to compensate for the increased sparseness we have marginally reduced the value of $\delta(G_r)$. Likewise, the $CC(G_r)$ we choose lies midway in the range (0.1 - 0.8) where the ends of the spectrum represent highly sparse and highly dense networks, respectively. Further, the $D(G_r)$ and $R(G_r)$ values demonstrate an even placement of nodes across the RWMN.\\
We now subject the above RWMN to the interference estimation schemes described earlier to assess the performance of CA schemes in the RWMN.
\begin{table} [h!]
\caption{Global Parameters' Values for the RWMN}
\centering
\tabcolsep=0.10cm
\begin{tabular}{|M{1.5cm}|M{1.8cm}|M{3.5cm}|M{1.3cm}|}
\hline 
\multicolumn{1}{|c|}{\textbf{Global}}&\multicolumn{2}{|c|}{\textbf{Range of Values}}&\multicolumn{1}{|c|}{\textbf{RWMN}}\\ \cline{2-3}
\multicolumn{1}{|c|}{\textbf{Parameter}}&\textbf{Full Range}&\textbf{Real-World Networks}&\textbf{Value}\\
\hline  
$\delta(G_r)$&0-1&0.09 - 0.1&0.083\\
\hline 
$CC(G_r)$&0-1&0.1 - 0.8&0.37\\
\hline  
$D(G_r$)&NA&NA&1649.88\\
\hline 
$R(G_r$)&NA&NA&168.21\\
\hline 

\end{tabular} 
\label{comp}
\end{table}
\subsection{Simulation Parameters}
We carry out rigorous simulations in ns-3 \cite{NS-3} to observe the performance of CA schemes in the $50$ node RWMN where the number of IEEE 802.11n radios/node is $3$ and every radio has an isotropic range of $250m$. The important simulation parameters are listed in Table~\ref{Sim}. A datafile is sent from a source node to a destination node in every multi-hop TCP flow. For UDP flows, data of identical size is sent from source nodes to destination nodes through UDP packets transmitted at an interval of 50ms. Similar set of test scenarios are evaluated employing TCP and UDP transport layer protocols. The TCP and UDP simulations are carried out independently. We consider two sets of metrics to monitor CA performance. The metrics for TCP simulations are \textit{average aggregate network throughput} (Throughput) and \textit{disrupted flow count} (DFC). Throughput is computed by first recording the aggregate throughputs for the RWMN for a variety of test cases, and then computing their average to obtain a 
measure of network capacity. Disrupted flows are those which fail to deliver the complete datafile at its intended destination owing to failures in routing mechanisms caused by high interference in the WMN 
\cite{Manas}. Thus, DFC is a measure of the intensity of interference in the network. A better CA scheme should exhibit lower DFC which will demonstrate its efficiency in alleviating interference in the RWMN. The metrics observed in the UDP simulations are the \textit{average end-to-end delay} (EED) and \textit{average packet delivery ratio} (PDR). RTS/CTS is disabled in the UDP simulations to assess the maximal adverse impact of interference on WMN performance and the ability of CA schemes to mitigate it without any assistance from the underlying 802.11 wireless frame collision prevention mechanisms. 

\subsection{Test Scenarios}
Since multi-hop data transmissions are the hallmark of WMNs, we employ various combinations of multi-hop flows, simulated concurrently to design test cases that effectively generate interference scenarios at varying traffic loads. Given the wide span of the simulated RWMN environment we ensure that the source and destination pairs are separated by 3 to 11 hops over which the data has to be relayed. This condition allows us to simulate interference scenarios which are very close to those experienced in real-world wireless networks. We engineer 5 test cases which consist of the following number of concurrent multi-hop flows~: \\ \quad i)~8 \quad ii)~12 \quad iii)~16 \quad iv)~20 \quad  v)~24.\\
The same set of test cases are subjected to both TCP and UDP simulations. 

 \begin{table} [h!]
\caption{ns-3 Simulation Parameters}
\raggedright
\begin{tabular}{|M{4.5cm}|M{3.5cm}|}
\hline
\bfseries
 Parameter&\bfseries Value \\ [0.2ex]
 \hline
\hline
Orthogonal Channels Utilized & 4 (5 GHz)\\
\hline
Datafile Size & 1 MB  \\
\hline
802.11n PHY Datarate&54 Mbps  \\
\hline
TCP ns-3 model &BulkSendApplication\\
\hline
UDP ns-3 model &UdpClientServer\\
\hline
TCP Maximum Segment Size&1 KB   \\
\hline
UDP Packet Size &512 Bytes\\
\hline
MAC Fragmentation Threshold&2200 Bytes  \\
\hline
RTS/CTS (TCP) & Enabled  \\
\hline
RTS/CTS (UDP) & Disabled \\
\hline
Routing Protocol &OLSR    \\
\hline
Rate Control&Constant Rate   \\
\hline
\end{tabular}
\label{Sim}
\end{table}  

\subsection{Selection of CA Schemes} 
Selection and implementation of CA schemes is a very crucial aspect of this evaluation. Thus, we implement a set of graph theoretic CA schemes making use of the \textit{enhanced multi-radio multi-channel conflict graph} (EMMCG) proposed in \cite{Manas}. The CA set is diverse in terms of network performance, including CAs of different performance levels which is ideal for our testing exercise.
The CAs range from a low performance Breadth First Search CA (BFS) \cite{22Ramachandran}, to radio co-location aware Elevated Interference Zone Mitigation CA (EIZM) which exhibits high network capacity \cite{Manas2}. Other CAs which lie between these two are the Maximum Independent Set based CA (MIS) \cite{24Aizaz}, the Link Preserving Interference Minimization CA (LP) \cite{LP}, the Graph Edge Coloring based CA (EC) \cite{EC}, and the radio co-location aware Optimized Independent Set CA (OIS). 
We carry out the performance evaluation of the CAs quantitatively and refrain from subjecting their algorithmic disposition to a qualitative analysis. Thus, the scope of this evaluation is limited to obtaining results of the stated thorough simulations and using them as a benchmark to determine the reliability of CA performance prediction schemes.
\begin{figure}[htb!]
                \centering
                \includegraphics[width=7cm, height=5cm]{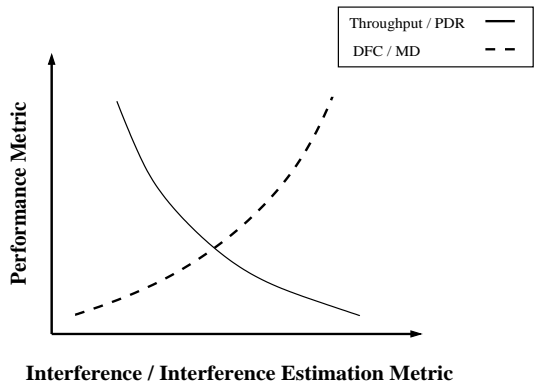}
                \caption{Expected correlation of network performance metrics with interference}
                \label{cor}
        \end{figure}
 \subsection{Results and Analysis}
\begin{figure*}
  \centering%
  \begin{tabular}{cc}
   \subfloat[TID vs Throughput]{\includegraphics[width=.33\linewidth]{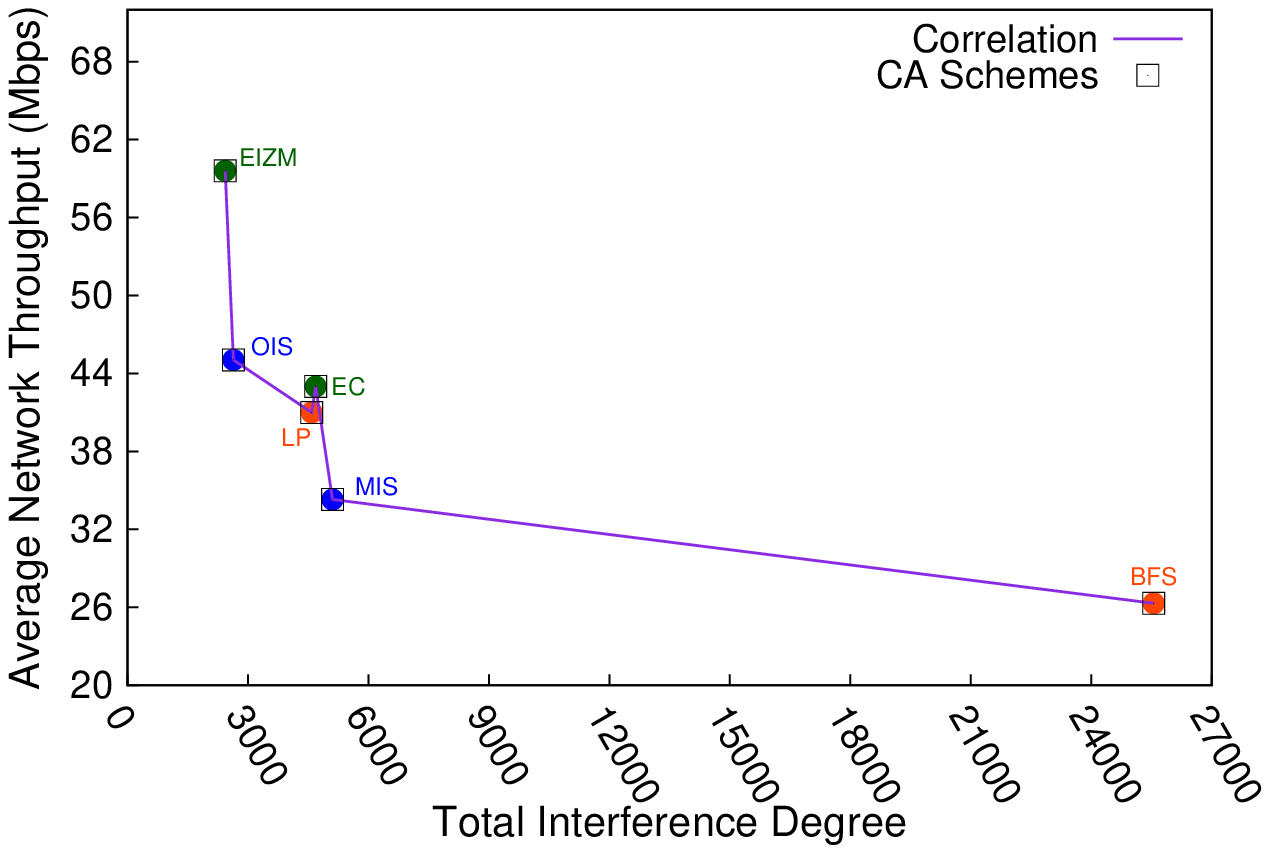}}\hfill%
    \subfloat[CDAL$_{cost}$ vs Throughput]{\includegraphics[width=.33\linewidth]{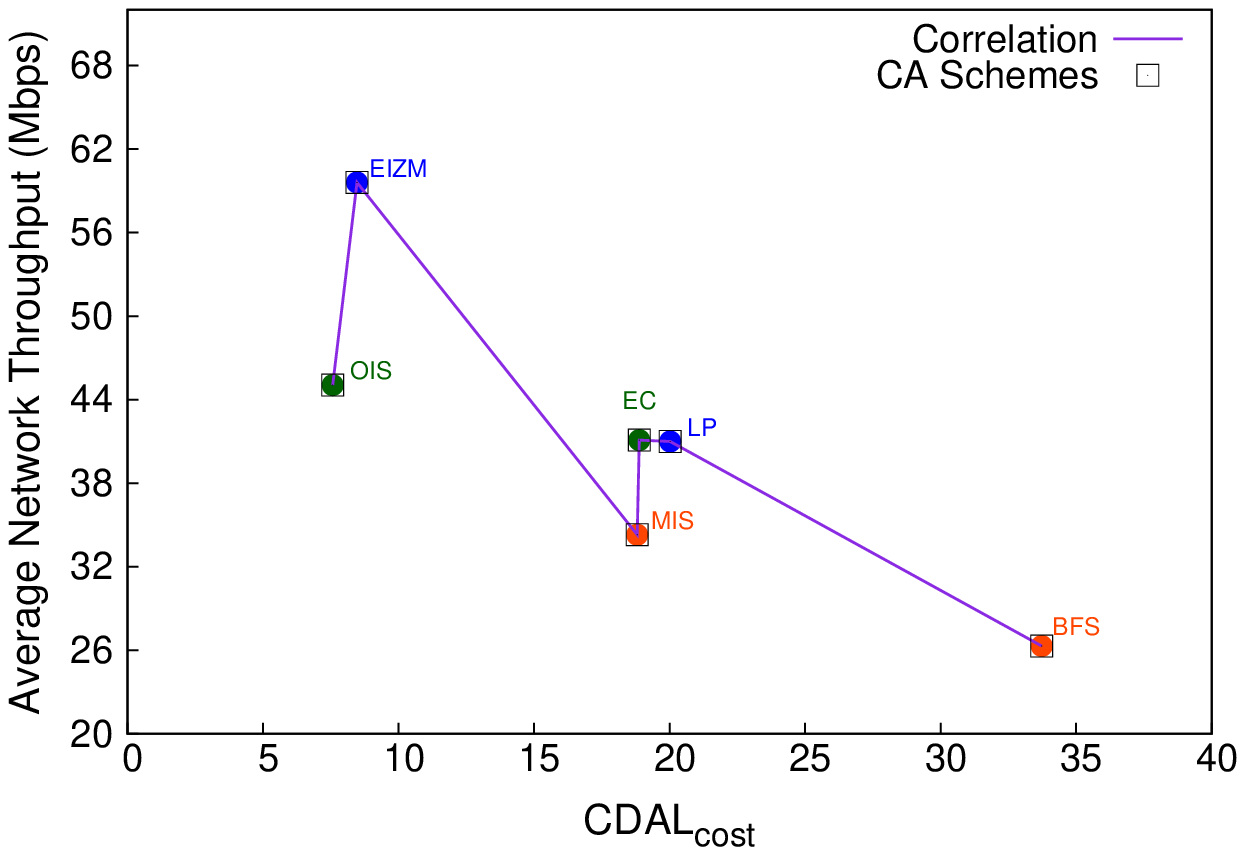}}\hfill%
   \subfloat[CXLS$_{wt}$ vs Throughput] {\includegraphics[width=.33\linewidth]{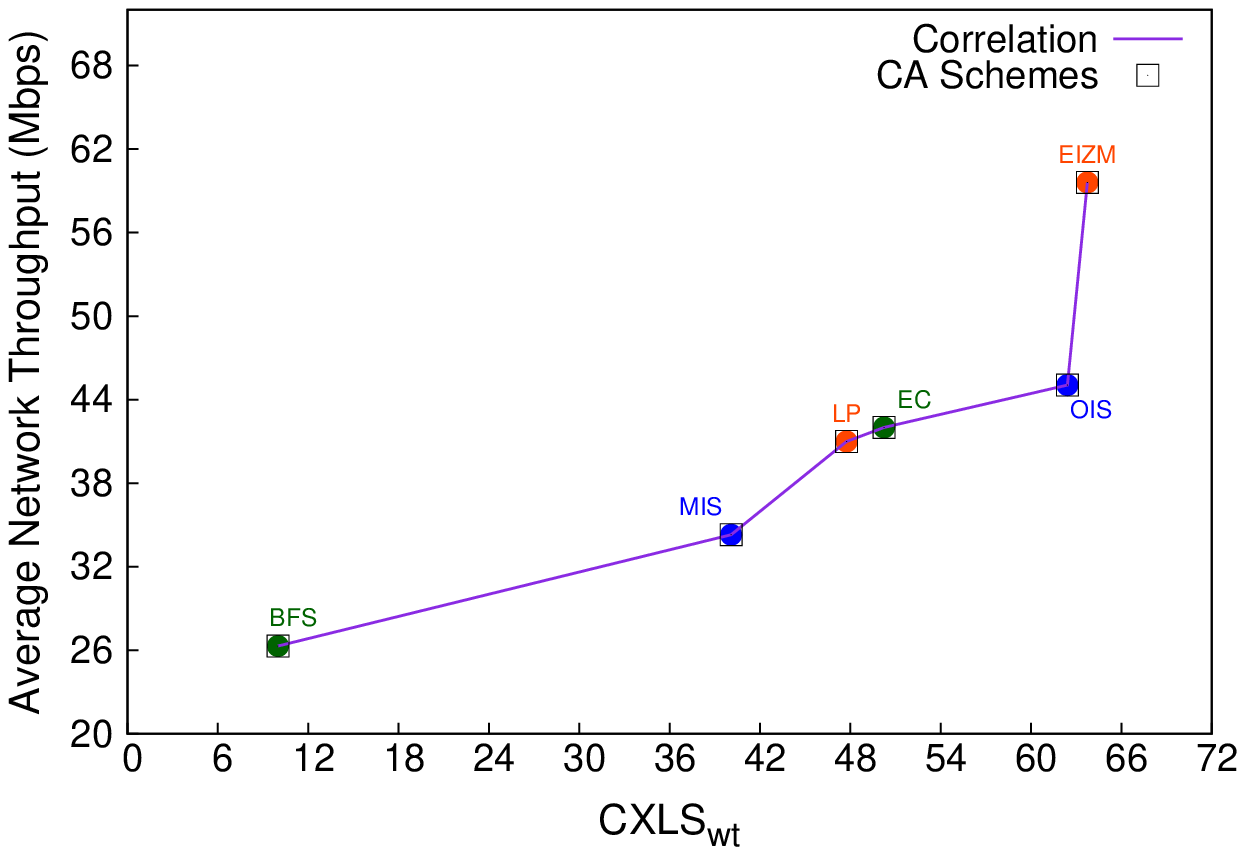}}%
    \end{tabular}
    \caption{Observed correlation of theoretical estimates \& Throughput} 
     \label{cTh}
\end{figure*}

\begin{figure*}
  \centering%
  \begin{tabular}{cc}
   \subfloat[TID vs DFC]{\includegraphics[width=.33\linewidth]{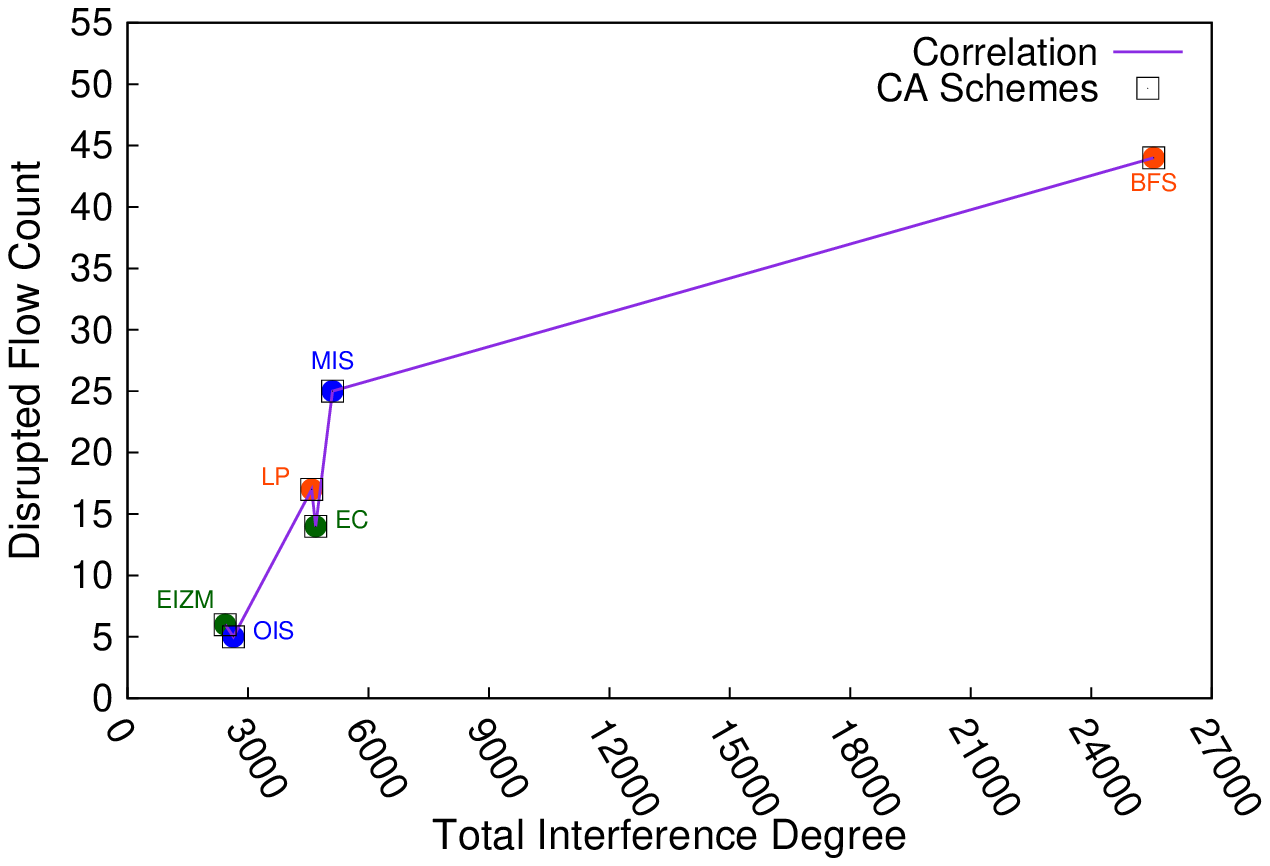}}\hfill%
    \subfloat[CDAL$_{cost}$ vs DFC]{\includegraphics[width=.33\linewidth]{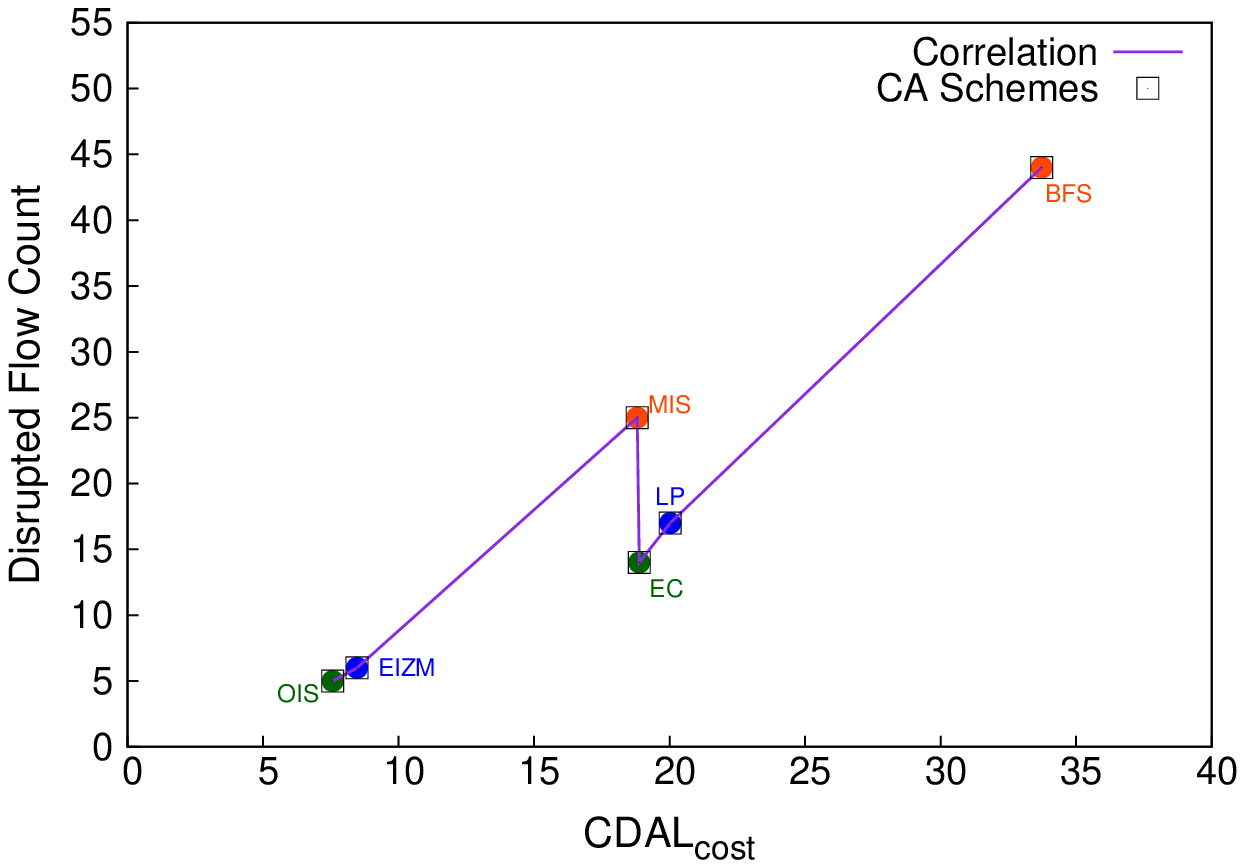}}\hfill%
   \subfloat[CXLS$_{wt}$ vs DFC] {\includegraphics[width=.33\linewidth]{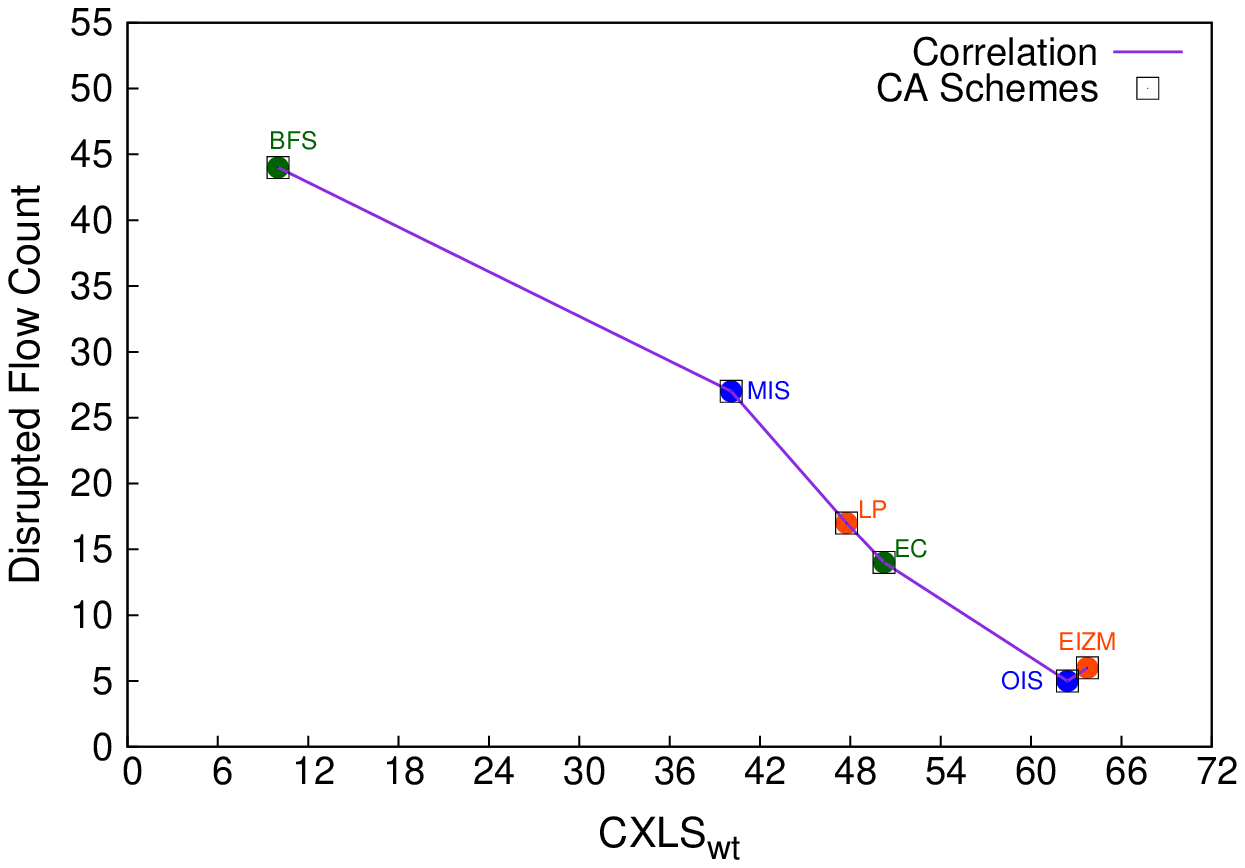}}%
    \end{tabular}
    \caption{Observed correlation of theoretical estimates \& Disrupted Flow Count} 
     \label{cDFC}
\end{figure*}

\begin{figure*}
  \centering%
  \begin{tabular}{cc}
   \subfloat[TID vs PDR]{\includegraphics[width=.33\linewidth]{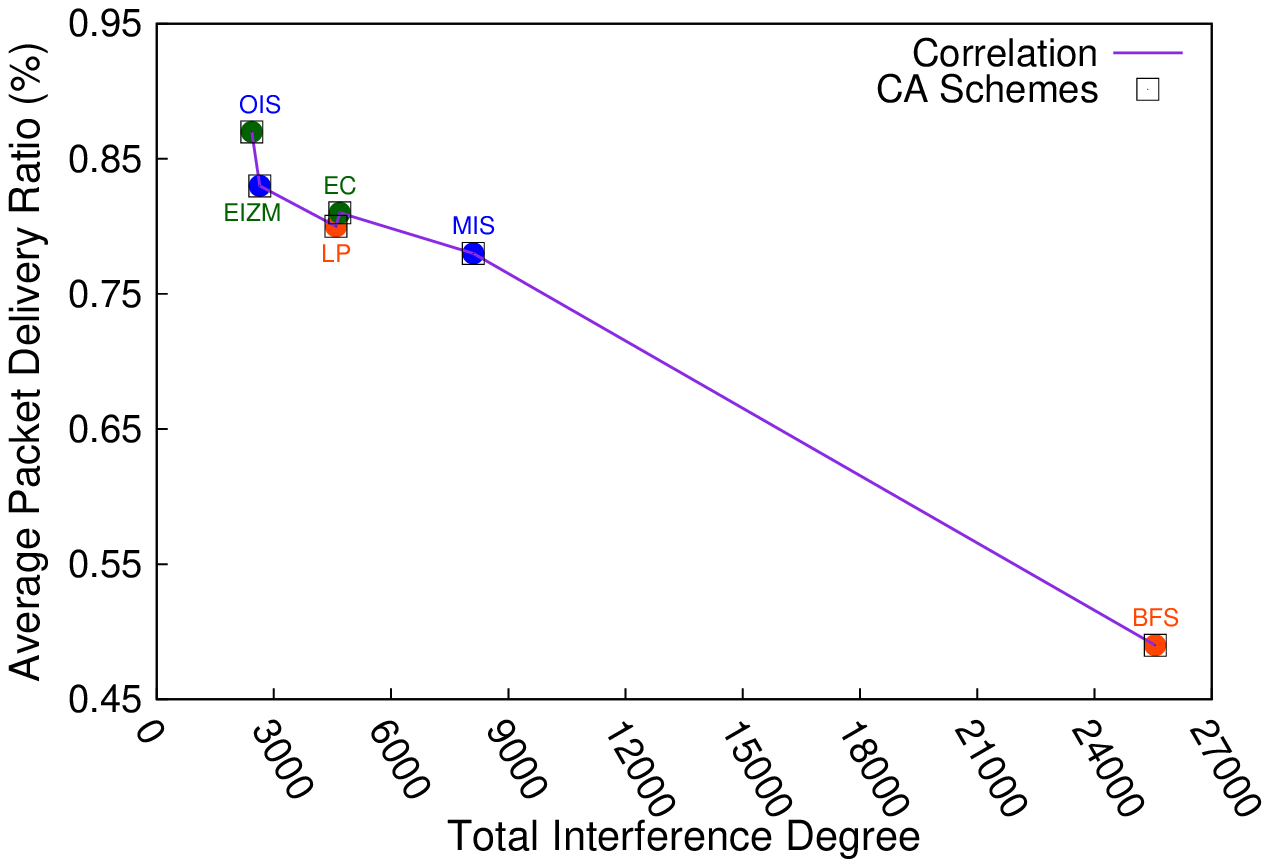}}\hfill%
    \subfloat[CDAL$_{cost}$ vs PDR]{\includegraphics[width=.33\linewidth]{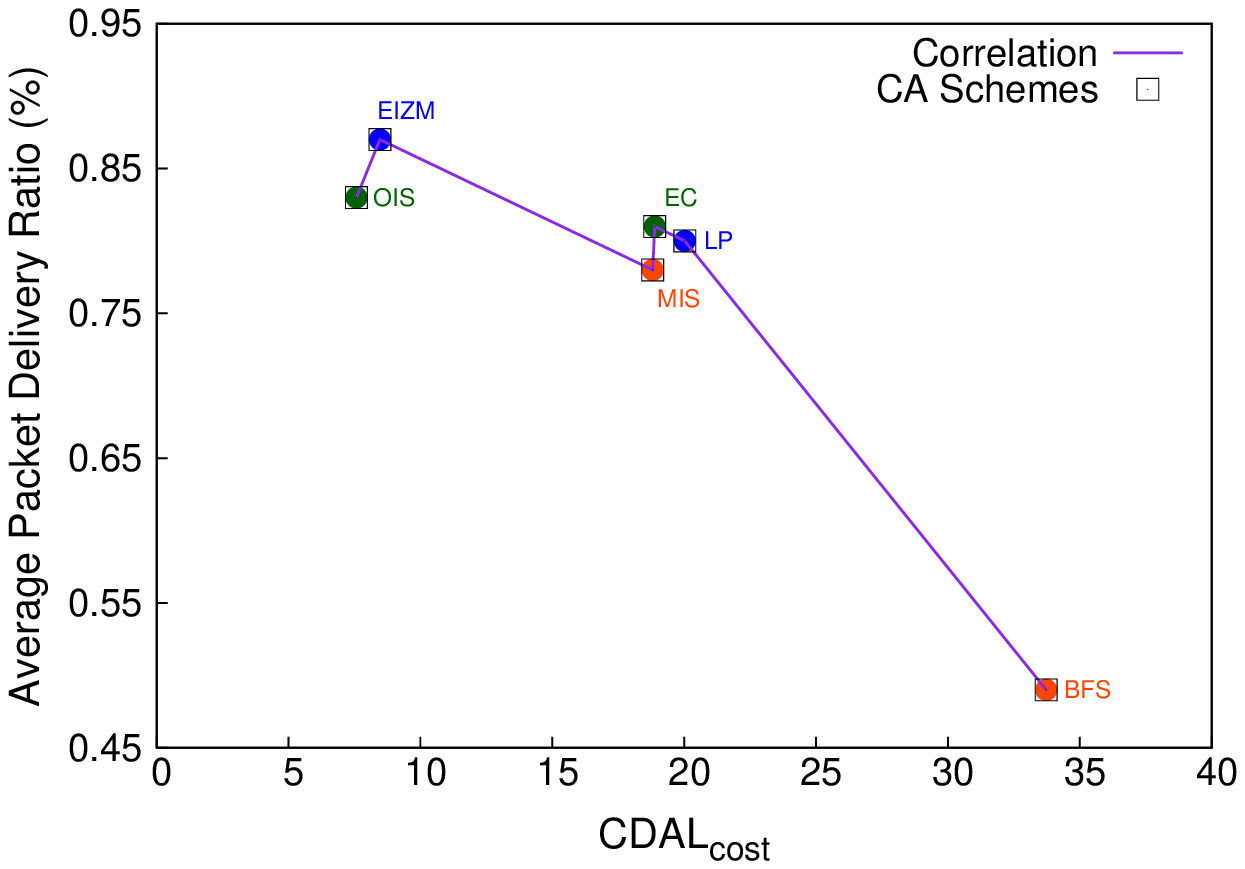}}\hfill%
   \subfloat[CXLS$_{wt}$ vs PDR] {\includegraphics[width=.33\linewidth]{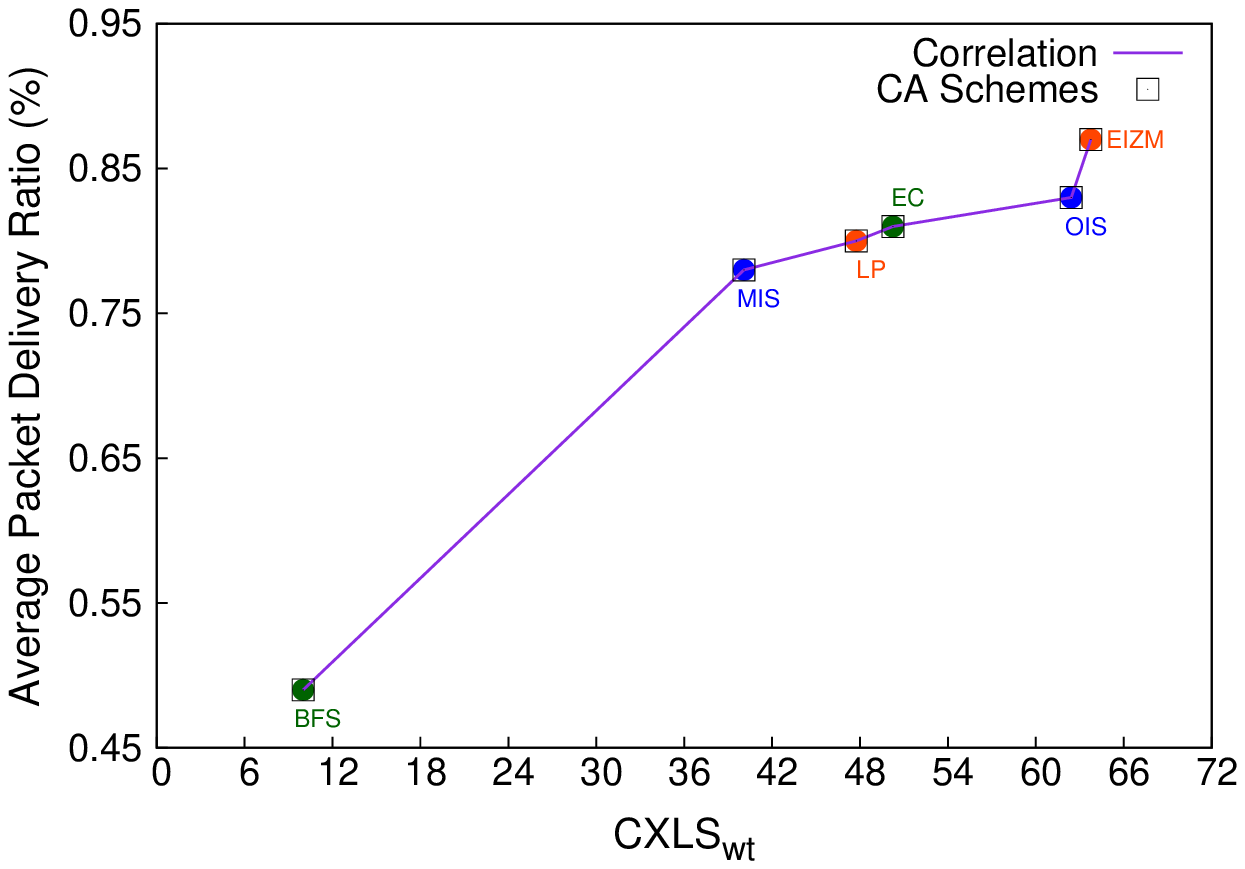}}%
    \end{tabular}
    \caption{Observed correlation of theoretical estimates \& Packet Delivery Ratio} 
     \label{cPLR}
\end{figure*}

\begin{figure*}
  \centering%
  \begin{tabular}{cc}
   \subfloat[TID vs EED]{\includegraphics[width=.33\linewidth]{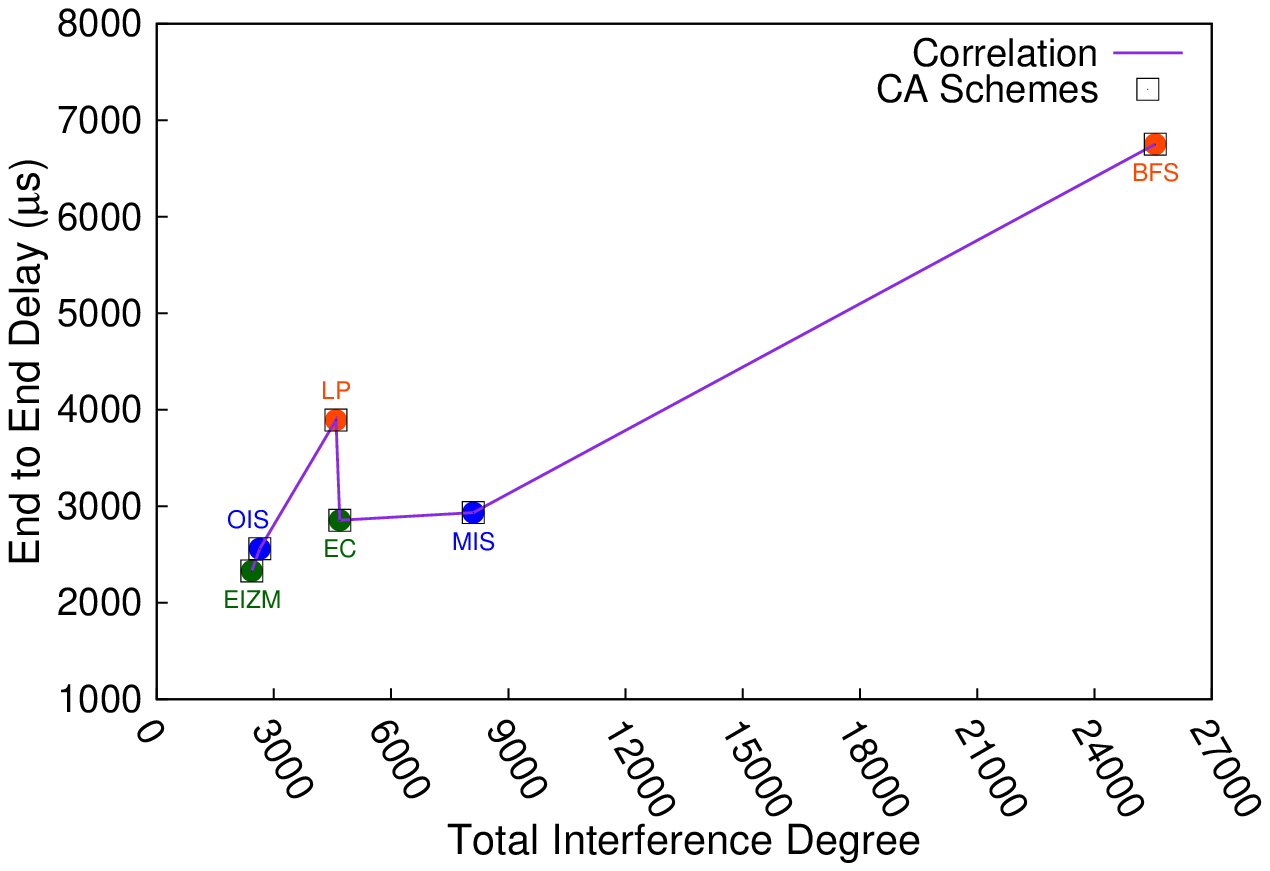}}\hfill%
    \subfloat[CDAL$_{cost}$ vs EED]{\includegraphics[width=.33\linewidth]{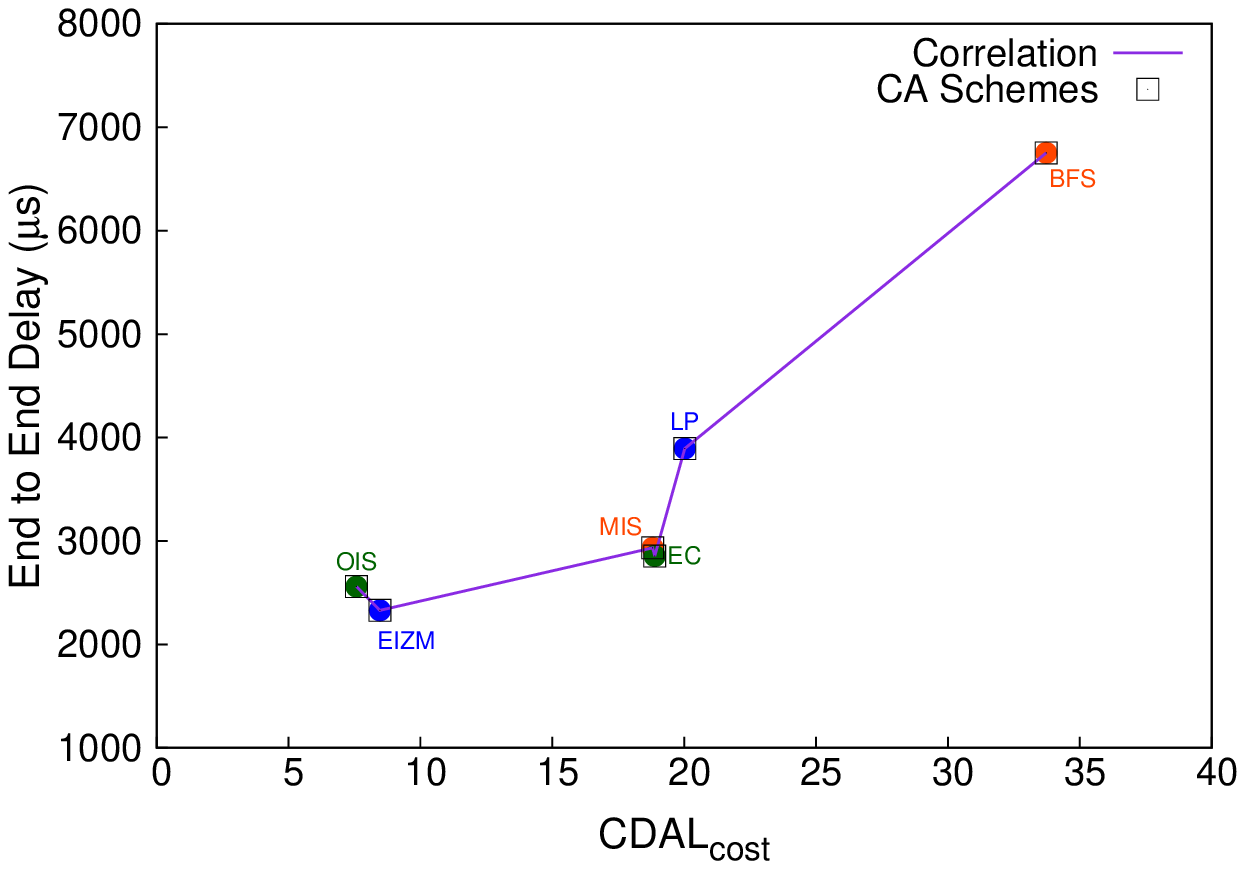}}\hfill%
   \subfloat[CXLS$_{wt}$ vs EED] {\includegraphics[width=.33\linewidth]{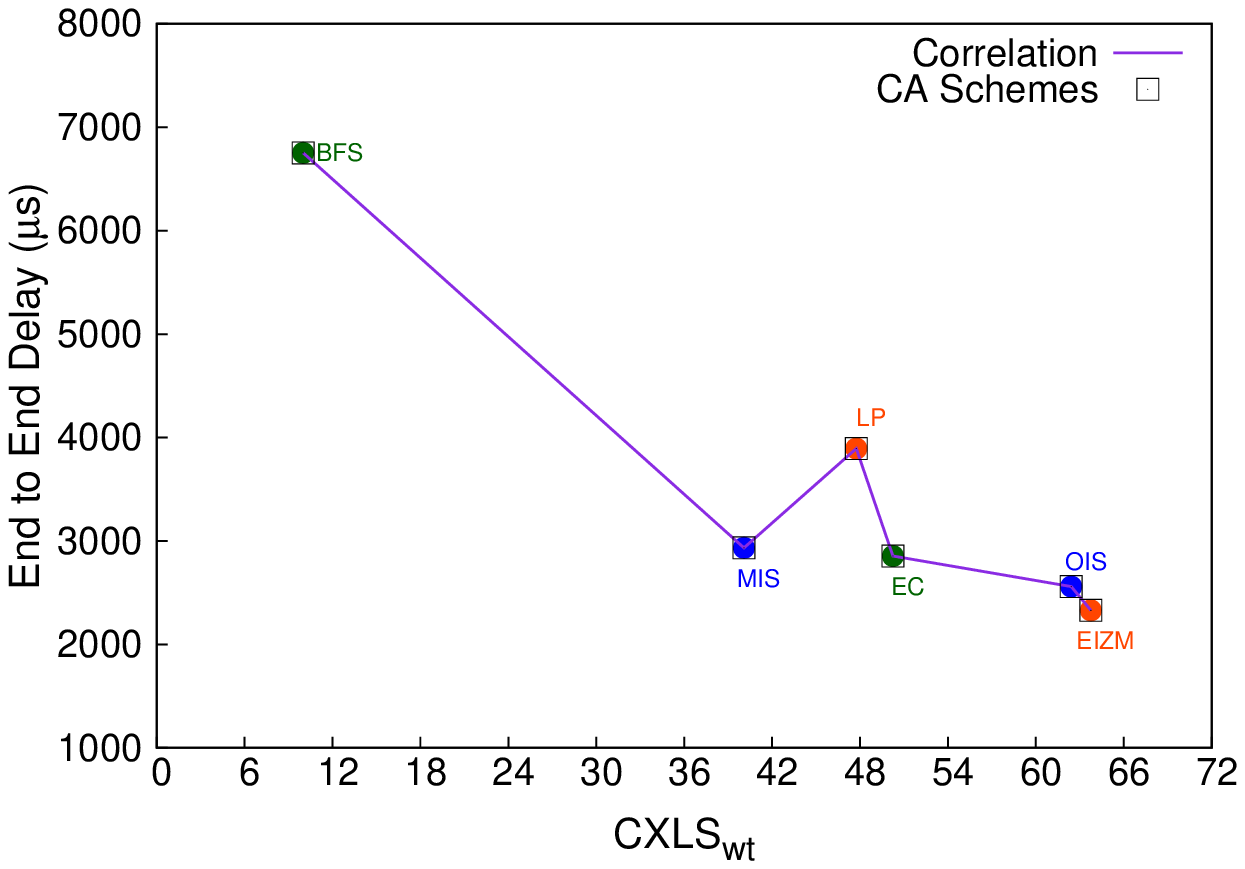}}%
    \end{tabular}
    \caption{Observed correlation of theoretical estimates \& End-to-End Delay} 
     \label{cMD}
\end{figure*}

\begin{figure}[htb!]
                \centering
                \includegraphics[width=8.5cm, height=5cm]{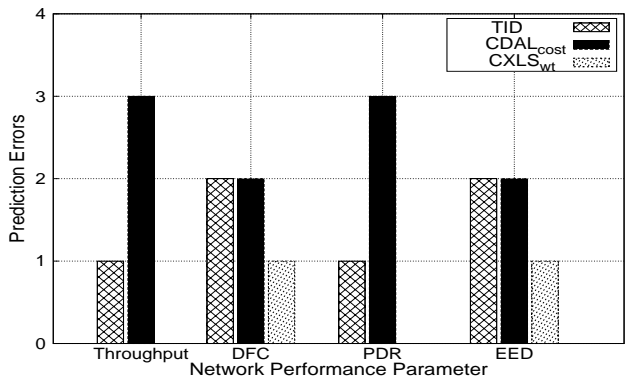}
                \caption{PE values of CA performance prediction metrics}
                \label{PE}
        \end{figure}

We record the values of the four network performance metrics (NPMs) being considered \emph{viz.}, Throughput (Mbps), DFC, EED ($\mu s$) and PDR (\%). Thereafter, we present the observed correlation of each of these NPMs with each of the three CA performance prediction metrics (CPPMs) \emph{viz.}, CDAL$_{cost}$,  CXLS$_{wt}$ and TID. To analyze the observed correlation it is imperative to understand the expected correlations between NPMs and the intensity of interference in a wireless network. We illustrate the expected correlation in Figure~\ref{cor}. With rise in interference the performance of a wireless network deteriorates. Thus, as the interference levels intensify the probability of attenuation of a carrier signal by an interfering signal in its proximity increases tremendously. This results in packet collisions and data loss, there by necessitating retransmissions of the corrupted data packets which add to the latency in successful packet delivery. Thus, a direct consequence of increased adverse 
impact of interference is decrease in PDR and an increase in EED. These in turn translate to a reduced overall network capacity or Throughput and a substantial increase in disruptions of data flows (DFC). The expected plots of these correlations are discernible from Figure~\ref{cor}, and we will use them as a benchmark to analyze and examine the plots of observed correlations.\\
We now present the observed correlation of Throughput, DFC, PDR and EED with each of the three CPPMs through Figures~\ref{cTh},~\ref{cDFC},~\ref{cPLR},~\&~\ref{cMD}, respectively. For ease of illustration we have labeled the CAs in the plots.
It can be inferred that among the three CPPMs, CDAL$_{cost}$ shows maximum deviation from the expected correlation with the NPMs while TID fares slightly better with a higher adherence. CXLS$_{wt}$ demonstrates highest compliance with the theoretical correlations exhibiting absolute conformity with respect to Throughput and PDR. In \cite{Manas4} where the simulation environment is a $5\times5$ grid WMN, TID exhibits the poorest adherence to expected correlations and CDAL$_{cost}$ fares much better. CXLS$_{wt}$ is the most reliable CPPM in grid WMNs as well.\\
We process the experimentally observed correlations to compute a measure of reliability for each of the CPPMs. We achieve this by establishing the \textit{performance relationship} between all CA pairs \emph{i.e.}, compare their performance in the RWMN. For all CA pairs we compare : a) The theoretical performance relationship predicted by the three CPPMs, with b) The experimentally observed performance relationship with respect to the four NPMs.\\
From these comparisons, we determine the \textit{prediction error} (PE) of each CPPM, which is the number of wrongly predicted performance relationships between CA pairs by a CPPM for a given NPM. The PE values of the CPPMs for each NPM are listed in Figure~\ref{PE}. Further, we compute the \textit{degree of confidence} (DoC) for each of the theoretical CPPM, which is a measure of accuracy of its predictions of CA performance. DoC value for a CPPM is computed through the expression $DoC = (1-(PE/\textsuperscript n C_2))\times100$, where $n$ is the number of CAs considered.  DoC for the three theoretical CPPMs are illustrated in Table~\ref{DOC}. 
\begin{table} [h!]
\caption{Performance Evaluation of Estimation Metrics}
\centering
\tabcolsep=0.10cm
\begin{tabular}{|M{2.5cm}|M{1.7cm}|M{1.7cm}|M{1.7cm}|}
\hline 
\multicolumn{1}{|c|}{\textbf{Performance}}&\multicolumn{3}{|c|}{\textbf{Degree of Confidence} ($\%$)}\\ \cline{2-4}
\multicolumn{1}{|c|}{\textbf{Metric}}&\textbf{TID}&\textbf{CDAL$_{cost}$}&\textbf{CXLS$_{wt}$}\\
\hline  
Throughput&93.33&80&100\\
\hline 
DFC&86.67&86.67&93.33\\
\hline 
PDR&93.33&80&100\\
\hline  
EED&86.67&86.67&93.33\\
\hline  
\end{tabular} 
\label{DOC}
\end{table}



\section{Conclusions}
CXLS$_{wt}$ offers the least prediction errors with respect to all the four NPMS \emph{i.e.}, Throughput, DFC, PDR and EED. The same pattern is reflected in the DoC values as well. TID estimate performs better than CDAL$_{cost}$ overall, but TID is decidedly a poorer CPPM than CXLS$_{wt}$. Although CDAL$_{cost}$ registers an average performance as a CPPM in the grid WMN layouts, it does not fare well in the RWMN with DoC values always below 90\% which implies poor accuracy in CA performance prediction. CXLS$_{wt}$ is the most reliable CA performance prediction metric among the three, with DoC values always greater than 90\%. Further, it outperforms the other two interference metrics in both grid and random layouts. Thus, CXLS$_{wt}$ is a topology agnostic CA performance prediction metric and offers reliable interference estimation.   
\section*{Acknowledgement} This work was supported by the Deity, Govt of India (Grant No. 13(6)/2010CC\&BT).

\bibliography{ref}

\end{document}